# Laser-driven Ion and Neutron Sources from Medium Repetition Ultrashort PW Laser

X. Jiao[1], C. Jeon[2], G. Tiwari[1,3], S. G. Lee[2], O. Labun[1], I. W. Choi[2], L. A. Labun[1,3], Mara Klebonas[1], C. Hojbota[1,2], D. D. Phan[1], C. H. Nam[2], B. M. Hegelich[1,2]

*1. Center for High Energy Density Science (CHEDS), Department of Physics, University of Texas at Austin, Austin, Texas 78712, USA*

*2. Center for Relativistic Laser Science (CoReLS), Institute for Basic Science (IBS), Gwangju 61005, Republic of Korea*

*3. Tau Systems, Austin, Taxes 78701, USA*

We report the first experiment investigating ion acceleration and neutron generation irradiating thin plastic targets ($CH_2$) and deuterated plastic targets ($CD_2$) of thickness ranging from 30nm to 160nm using the 4 PW (0.1 Hz) laser at CoReLS in South Korea. Thin wedge-shaped filters exploiting differing stopping ranges were designed to distinguish carbon 6+ ions from deuterons in shots with $CD_2$ targets. The maximum energies of all ion species from both $CH_2$ and $CD_2$ targets were found to increase linearly with the laser intensity. The highest observed energy of each ion species scales as q (charge)$^2$/mass, which is more similar to the scaling expected for ponderomotive acceleration than to the scaling expected for TNSA. The maximum ion energies were also found to increase with target thickness. Utilizing the secondary interactions of the deuteron beam, we created a fast neutron source via deuteron breakup reactions on a copper converter. The neutron spectrum follows an exponential distribution with energy up to ~15 MeV. A neutron yield of ~$2\times10^7$ n/J was observed from thicker targets, comparable to TNSA-regime laser-driven neutron sources and within one order of magnitude of the highest yields reported using Break-out-afterburner (BOA) acceleration with beryllium converters.

## Introduction

Over the past two decades, ion sources from laser-solid interactions and ion-driven neutron sources have attracted significant interest due to their unique characteristics and wide applications. Laser-plasma acceleration offers alternative means to generate fast particle beams in distances much shorter than radio frequency particle accelerators. Short-pulse, high-power laser accelerated ions are energetic enough to trigger nuclear reactions resulting in neutron production, opening new avenues to generate fast neutrons [12][13]. These neutron sources exhibit shorter temporal structure and higher peak flux compared to traditional spallation or reactor neutron sources. They are also compact in terms of both facility scale and source size. The high neutron flux produced through these sources is favored by applications such as neutron resonance radiography and neutron resonance spectroscopy.

Most neutron generation strategies employ a “pitcher-catcher” configuration, in which the ions accelerated from the primary target (pitcher) bombard a secondary target, also called converter (catcher), placed after the primary target. Nuclear reactions in the converter can lead to a short burst of neutron emission, whose energy spectrum and yield depend on the species and energy spectrum of the bombarding ions and the material of the converter. Optimizing the neutron yield or energy thus reduces to optimizing the ion spectrum from the laser-driven plasma in the primary "pitcher" target.

In a typical laser-thin foil interaction, ions are accelerated by the electric fields created by the charge separation because of heated electrons escaping from the target through the rear surface. This mechanism is referred to as target normal sheath acceleration (TNSA). A typical TNSA mechanism involves the interaction of a linearly polarized, ultrashort and ultra-intense laser pulse with solid foil targets ranging in thickness from sub-micron to several tens of microns [1][2]. Several optimization efforts including target engineering and laser pulse parameter control have been conducted to improve this technique. Although the mechanism is very robust, it favors the acceleration of the lightest ions, mainly protons, due to their highest charge-to-mass ratio and intrinsically produces continuous, exponential spectrum with broadband energy and wide angular divergence [3].

Alternative ion acceleration schemes were suggested to improve ion beam quality. One method utilizes the radiation-pressure of the laser pulses to drive the acceleration, called radiation pressure acceleration (RPA) [4]. It is also referred to as light-sail acceleration when ultra-thin targets are used. The laser pulse drives the electrons of the over-dense plasma forward volumetrically, leaving behind ions and creating a charge separation, which in turn accelerates the ions. Since the target advances, all ions are accelerated to similar speeds in principle. However, RPA is very demanding on laser temporal contrast because thin targets and high laser intensities are desired for this mechanism to dominate others[5]. For thicker targets, the intense radiation pressure can only drive electrons at the surface inward. This results in a parabolic deformation and laser penetration of the target while creating large charge separation. This process is called hole-boring (HB) [6]. If the electrons gain relativistic energies in the ultra-intense laser field, the relativistically-corrected plasma frequency can drop below the laser frequency, making the target transparent to the laser field. This enables break-out after burner (BOA) and has been reported to lead to enhanced ion acceleration through several interaction stages[7][8].

In the last two decades, multiple experiments have used high power laser facilities to create pulsed neutrons by accelerating and passing ion beams through low-Z converters, including deuterium, beryllium-9, and lithium-7 [11]. However, few experiments have been conducted using high-Z converters. Although these systems can generate neutron sources with good energy efficiency, all of them operate at a very low repetition rate. This limits their usability in real applications. On the other hand, multi-PW femtosecond Ti:Sapphire lasers can achieve the same focused laser intensity with less pulse energy and is capable of performing at the repetition rates up to 0.1 Hz at high intensities [9][10]. In this paper, we report the first laser and ions driven neutron experiment conducted on such a laser system at the Center of Relativistic Laser Science (CoReLS), South Korea. This experiment was set to explore the factors that affect ion acceleration and neutron generation under such conditions; we also studied the relationship between the laser intensity, ion energy and neutron energy and yield.

## Experimental setup

This campaign consisted of two stages. In the first stage, we systematically studied ion acceleration by irradiating ultra-thin plastic ($CH_2$) targets and deuterated plastic ($CD_2$) targets ranging from 30 nm to 160 nm (see Fig. 1 (a)). In the second stage, we directed the accelerated deuteron beams from $CD_2$ targets into a copper converter to create a neutron source, as shown in Fig. 1b. The experiments were carried out with the 0.1 Hz, 20 fs, 4 PW Ti:sapphire laser at CoReLS, where the temporal contrast exceeded $10^{10}$ after employing the double plasma mirror configuration[reference]. In the campaign, we used an F/1.5 off-axis parabolic mirror (OAP) with close to normal incidence angle to focus the linearly polarized laser light into a ~1.8 μm diameter (FWHM) spot with 40% to 50% enclosed energy. Starting with pulse energy up to 90 J before compression, after accounting for losses through the grating compressor [14] and the double plasma mirror system (total reflectivity 70%) [15], the pulse energy at the target plane was estimated to be up to ~30-50J, yielding peak intensities at the geometrical focus of F/1.5 OAPM of up to $5 \times 10^{22}$W/cm². The corresponding measurements of laser pulse duration and beam parameters were recorded for each shot during the campaign.

The in-house built $CH_2$ targets were used to accelerate proton and carbon ions at various charge states. The $CD_2$ targets (98% D), fabricated by Institut für Kernphysik in Germany, were used to accelerate deuteron ions. These targets were first spin coated on a silicon wafer and then floated onto square-shaped metal target frames containing 32 holes to avoid unevenness and wrinkles on the targets. A motorized target mount capable of holding 9 frames simultaneously was used, providing a total capacity of ~270 shots per mount load. The target frames were mounted at a tilt angle of 15° with respect to the laser beam axis to prevent specular back-reflection into the laser chain. The frames ensured consistency between shots, since the laser plasma interaction is very sensitive to the target surface condition. The mount allowed very rapid replenishment of targets when used with a high repetition rate laser, which enabled us to take hundreds of laser shots a day when the laser condition permitted and draw statistically reliable conclusions.

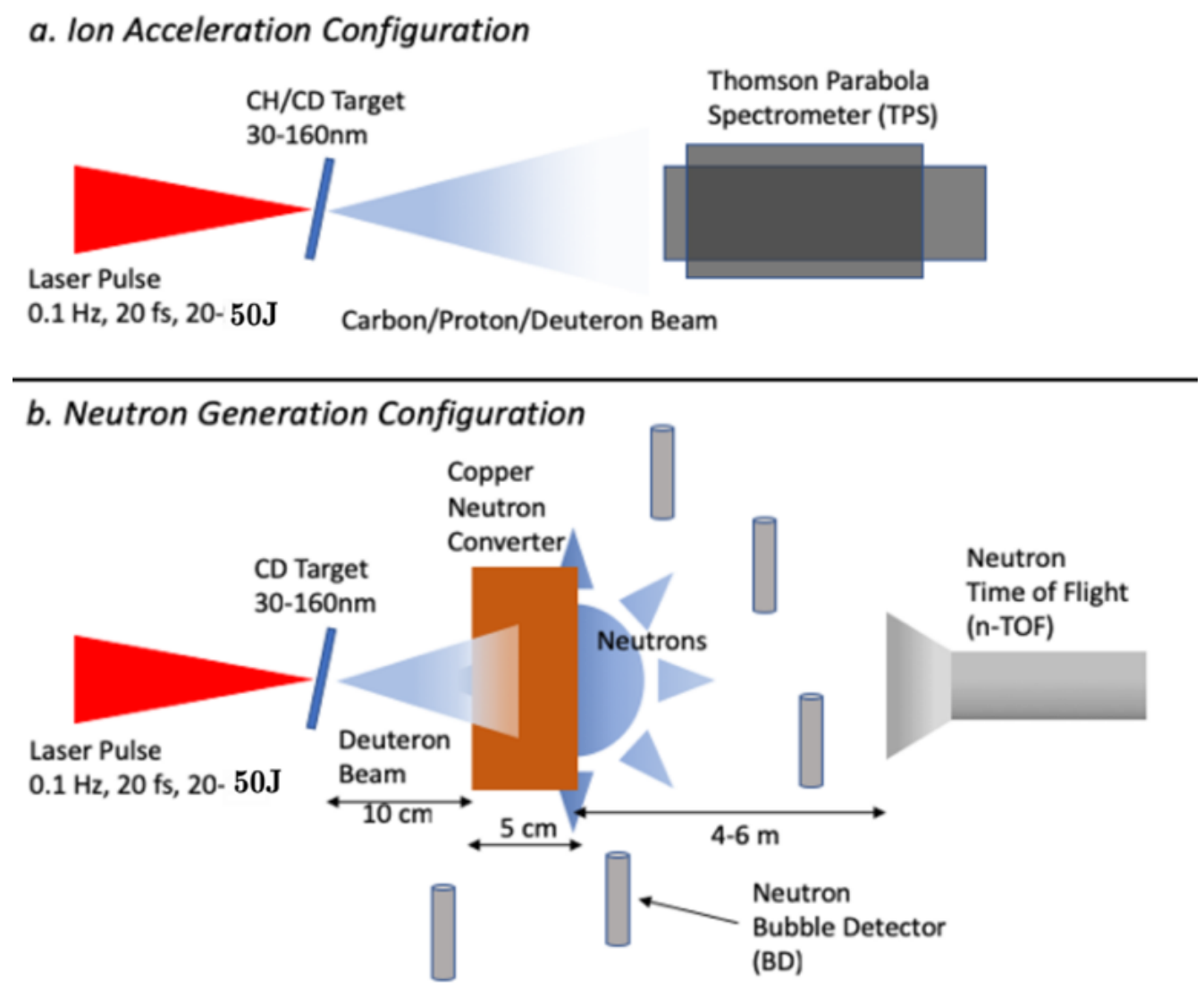

Figure 1. Illustration of Experimental Configurations.

We used three Thomson parabola spectrometers (TPSs), each with a 400 μm diameter pinhole, aligned horizontally just outside the target chamber, ~110 cm from the target, to measure ion spectra. These TPSs were positioned horizontally at the angles of 0°, 30° and -30° with respect to the laser propagation direction. One drawback of TPSs is that they can only distinguish ions with different charge to mass ratio (q/m). For instance, the traces of carbon 6+ ions and deuterons will overlap because they have the same charge to mass ratio, making it difficult to isolate the deuteron energy spectrum from $CD_2$ targets.

For this reason, we designed and implemented wedge-shaped filters exploiting the difference in stopping ranges between these two ions to stop carbon ions but let deuterons pass through. For non-relativistic momenta, carbon ions have kinetic energy 6 times greater than deuterons found at the same spot on the detection plane in a TPS. The stopping power [16] of an ion in matter is proportional to q^2 as opposed to the linear relation between ion energy and its charge. To design the wedge filter, we first calculated calibration curves of the TPSs, shown in Fig. 2a, where x-axis is the displacement on the TPS screen and the y-axis is the corresponding ion energy at that location. Then, the stopping range of the ions at each location is calculated based on the continuous slowing model with the stopping power given by the Bethe formula and shown in Fig. 2b, where the y-axis is the stopping range of the ions at each location. We chose aluminum as the filter material for its low electron density which allows a longer stopping range and a larger difference in the stopping ranges for different ions. The thickness profile is chosen so that the wedge filter not only differentiates these two types of ions but also attenuates x-rays to suppress their contamination of the deuteron trace.

Fig 2c and 2d show typical TPS images with and without the wedge filter installed. The vertical line at “zero point” in Fig 2c is a mark left by a fiducial wire on the microchannel plate (MCP) screen of the TPS and was used to identify the zero position on the filter. This metal wire aids the alignment of the TPS and monitors the position of the filter during the laser shots by blocking x-rays from the laser-target interaction. We found that the mark of the fiducial coincided exactly with the zero point of the TPS, which was determined by shooting targets with very low power pulses to produce a tiny unsaturated dot on the screen. In designing the wedge filter, we extended the filter along the x-axis direction on purpose to block an interval of the deuteron trace as shown in Fig 2b and 2c. Knowing the energies corresponding to the ends of this interval, we verify the TPS calibration and identify the impact on the MCP screen of x-rays that are generated by the ion-matter interaction.

The TPS images are plotted in log scales. A gap is visible in the trace where all ions are blocked, and no x-ray signal observed. Since this is the thinnest part of the filter, we can safely assume that x-rays generated at the other, thicker side of the filter neither leak to the MCP screen nor affect the trace signal. In addition, the deuteron ions traveling through the filter will slightly deviate from their original trajectories due to scattering, creating a more faint and blurred trace. Although we cannot recover the deuteron spectra without knowing how much the signal has been reduced by the filter, the wedge filter still enables us to identify the highest deuteron energy accelerated from the $CD_2$ targets. We ensure consistent comparison by using the same MCP signal level to determine the highest energy from all laser energy and target thickness combinations.

Neutrons were generated from a 5 cm thick copper converter about 10 cm behind the $CD_2$ target via the breakup reaction $d \rightarrow p + n$. For absolute yield measurements, we used 20 bubble detectors (BD-PND) with sensitivities of ~20 bubbles/mrem. These bubble detectors have roughly flat response to neutrons in the energy range 300 keV to 30 MeV and are insensitive to x-rays. The bubble detectors utilize super-heated fluid droplets that are stable in a polymer matrix at normal room temperature. The fluid is rich in protons and forms visible bubbles when energy is deposited by the passing neutrons via a (n, p) recoil reaction [18]. The detectors were previously calibrated using a radioactive neutron source and were within shelf life (3 months) during the campaign. The neutron spectrum in different directions was measured by three SCIONIX neutron time-of-flight (n-TOF) detectors. The signals were recorded by a fast-digital oscilloscope (LeCroy WM816Zi). The n-TOFs were composed of 150 mm diameter, 25 mm thick fast decay plastic scintillators (EJ-228) coupled to fast photomultiplier tubes (Hamamatsu R2083). The n-TOF detectors were shielded from x-rays by a layer of 5 cm thick lead bricks. The distance from converter to n-TOFs ranged from 4.9 m to 10.4 m. Two of the n-TOFs were placed in the forward and the backward directions horizontally. One n-TOF was placed in the side direction overlooking the target chamber from above.

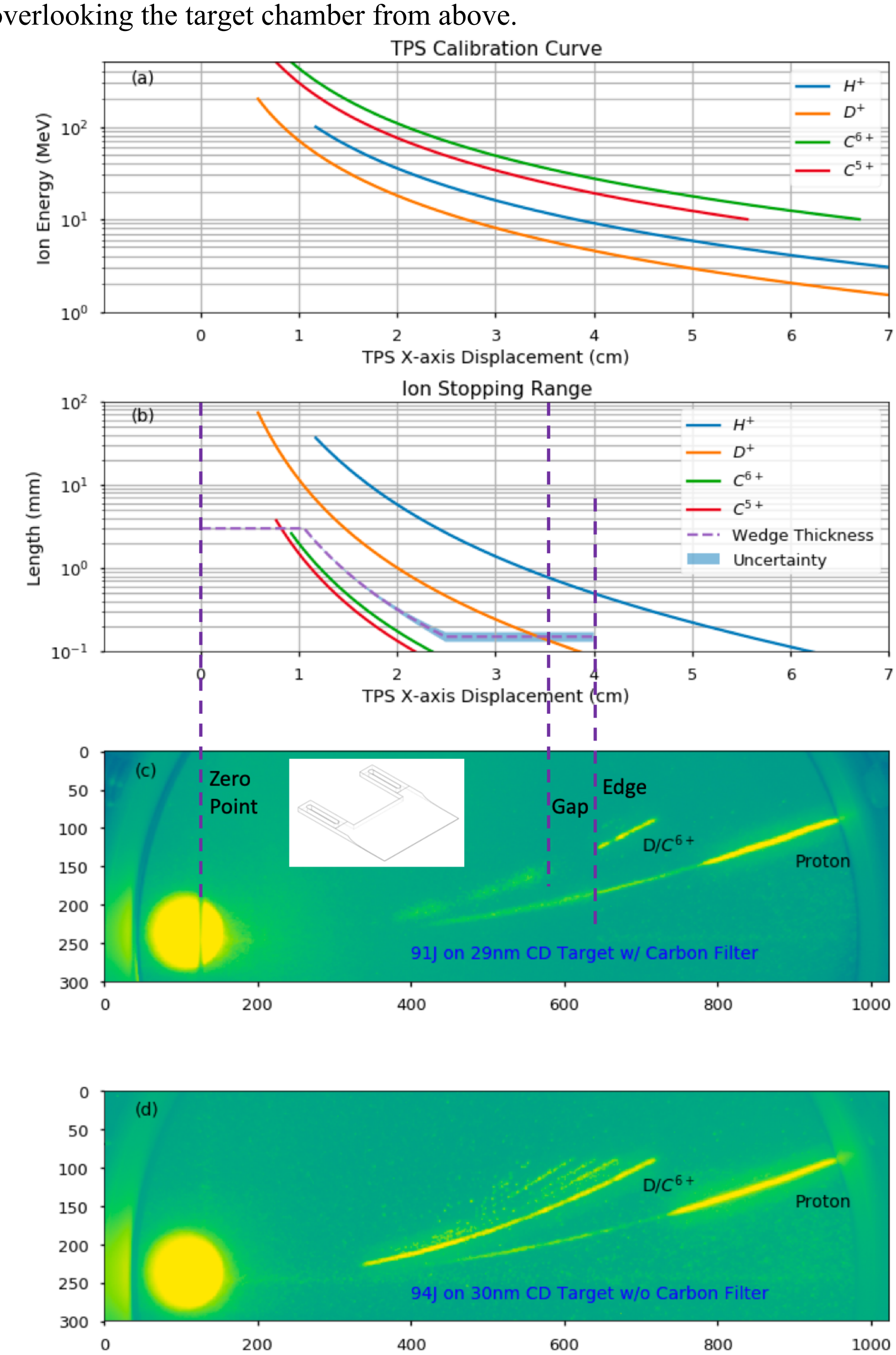

Figure 2 Calibration and design of wedge filter for TPS. (a) TPS calibration curves for primary ion species of interest. (b) Ion stopping range vs TPS position with wedge thickness profile overlaid. TPS raw signal from a laser shot with (c) and without (d) the wedge filter in place. In (c), the interval where no deuteron signal is expected coincides with the observed gap.

## Results

Fig. 3a and 3c show the energy spectra of carbon 6+ ions and protons measured in the laser propagation direction ($0^o$) from 100 nm $CH_2$ targets for various incident laser energy values. The plotted spectra present the signal after background subtraction, and they generally follow an exponential distribution. The x-axis is the ion energy in MeV and the y-axis is the signal level per MeV. A black line shown in Fig. 3a and 3c at $10^4$ count/MeV was taken as a reference threshold between signal and noise to determine the maximum ion energy, and implicitly our detection limit. It is evident that the signal becomes very noisy below this threshold. The scatter plots, Fig 3b and 3d on the right side, show trends in the summary statistics extracted from the spectra at left. The red dots present the maximum ion energy at each laser energy and correspond to the scale on the left y-axis. The blue dots, corresponding to the right y-axis, present the total energy fluence through the TPS pinhole and is calculated by summing the energy in the spectrum. The dashed lines represent least squared fits for each summary statistic.

The maximum energy for each ion species increases linearly with the laser intensity ($E_{ion} \propto E_L$). We only present the results from 100 nm $CH_2$ targets, but we encountered a linear relationship for all other thickness of $CH_2$ as well as for protons, C6+ and deuterons from $CD_2$ targets. The total energy fluence increases quadratically with input laser energy. The growth of the total energy fluence is due to the increase in the ion number as well as the hardening of the spectrum. At higher laser intensity, the population of particles in higher energy range increases faster than the that of particles in the lower energy range. For incident laser energies (>30 Joules), the ion energy spectral shape does not change much whereas the population of particles at all energies increases roughly proportionally.

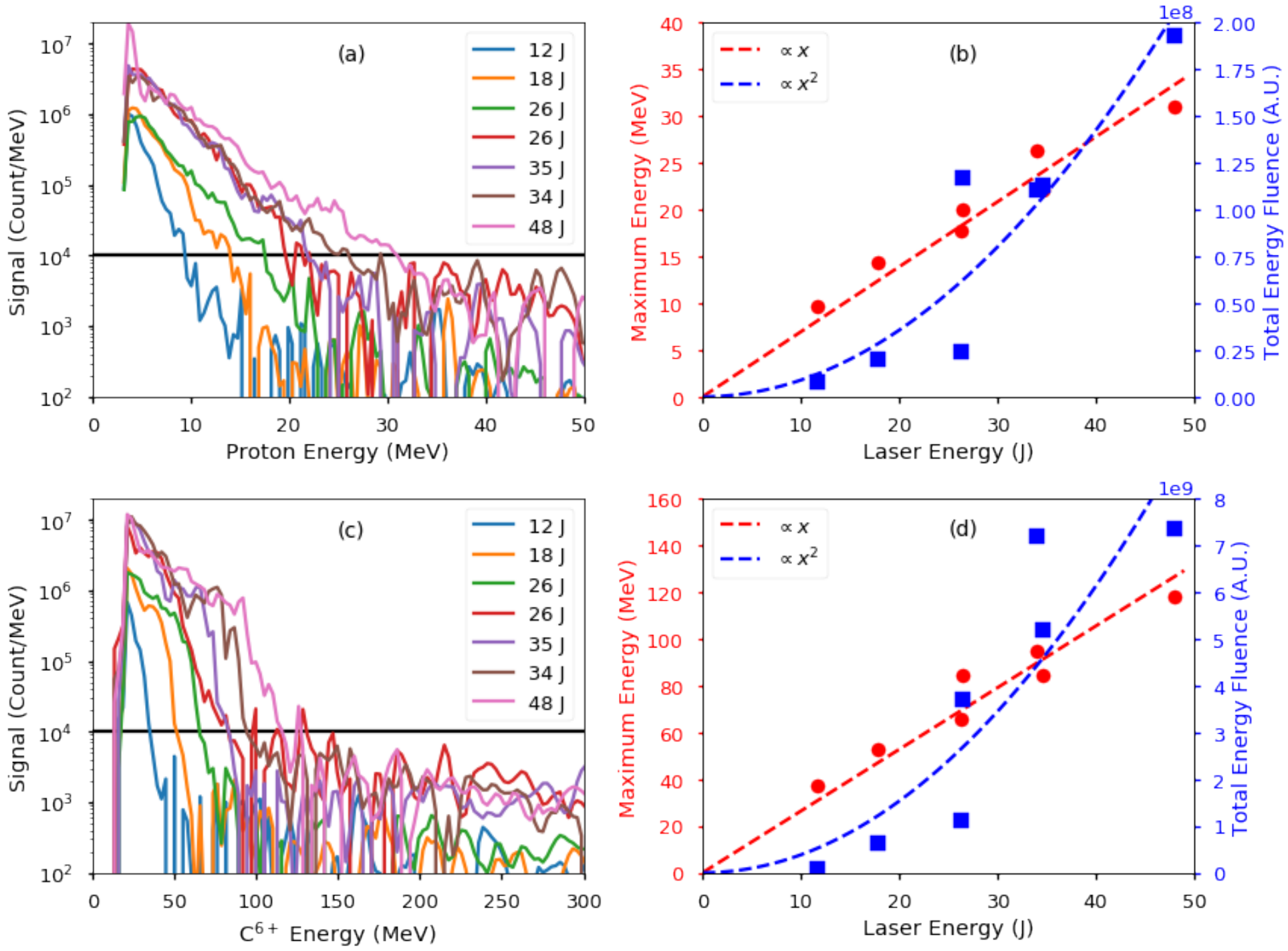

Figure 3 (a) Proton spectra for different laser energies. (b) C6+ spectra for different laser energies. (c) Laser energy dependence of maximum proton energy and total (integrated) energy in the spectrum with fits (dashed lines). (d) Same as (c) but for C6+.

Fig 4. compares the scaling law for maximum ion energy across all the identifiable ions in 100 nm $CH_2$ targets as well as in 101 nm $CD_2$ targets. In contrast to Fig 3, the maximum energy in Fig 4a and Fig 4b for each ion species has been rescaled by the ion's atomic number ($Z_{ion}$). We find that protons and carbon ions of different charge states all exhibit a linear relationship between maximum ion energy and laser energy. Least-square fits confirm that the proton has the highest attainable energy per ion charge per joule of laser energy, followed by carbon 6+ then carbon 5+, etc. The bar graphs on the right, Figs 4b and d, show the slope of the fits as a function of charge to mass ratio *q/m*. The linear trend in Figs 4b and d show that the maximum ion energy is proportional to $q^2/m$, where q is the ion charge and *m* is the ion mass. This scaling law is found to apply for all ions: protons, deuterons, and carbon ions in different charge states. The scaling law predicts for example that the maximum C6+ energy would be $6^2/5^2$ (1.44x) that of C5+ ions, 3x that of protons, and 6x that of deuterons in each laser shot.

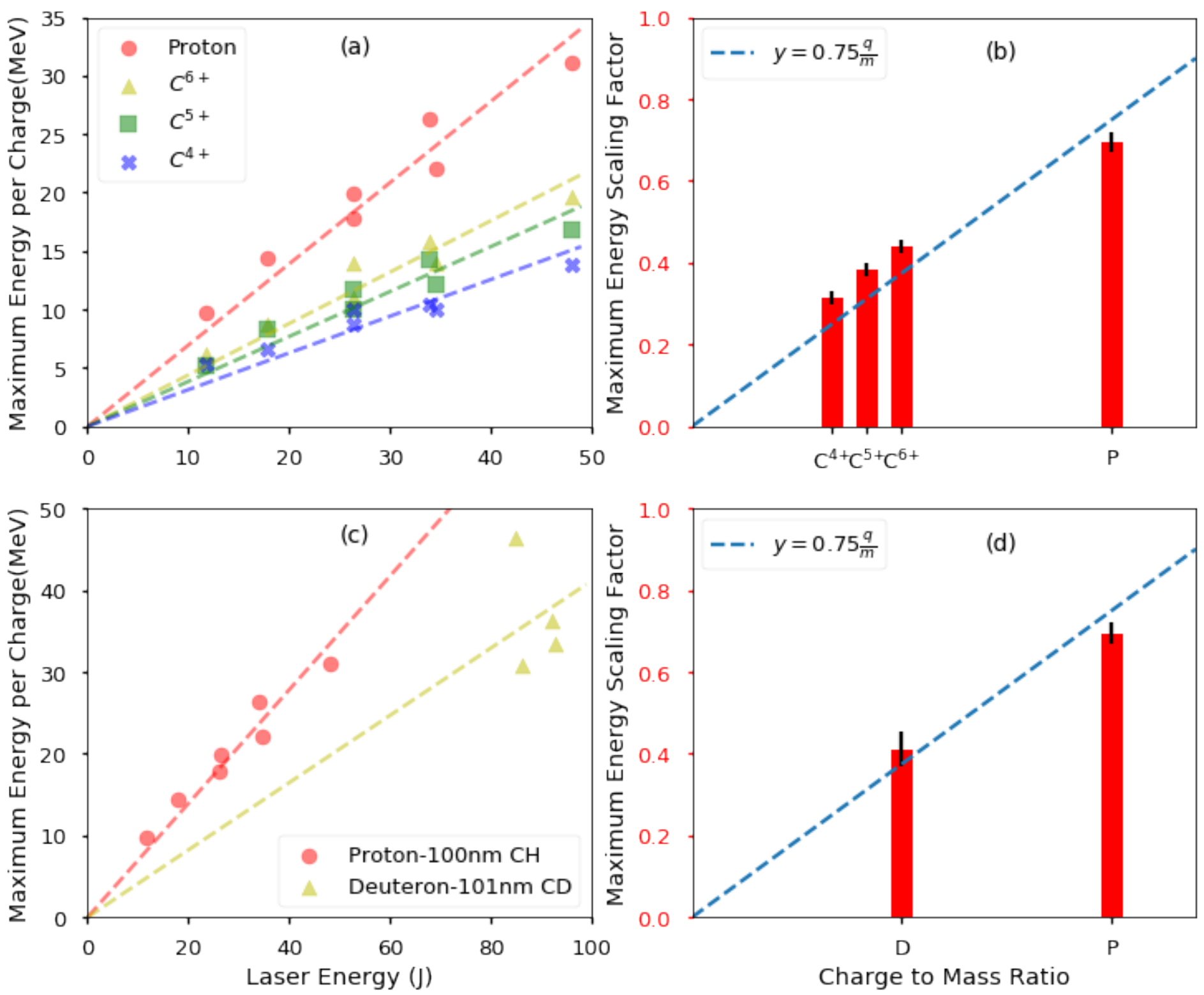

Figure 4 Observed scaling law for maximum ion energy. (a) Maximum energy divided by ion charge (Z) as a function of laser energy for CH2 targets. (b) Maximum energy divided by ion charge (Z) as a function of laser energy comparing protons from CH2 to deuterons from CD2. (c) Slope of the linear fit in (b) as a function of q/m. (d) Slope of the linear fit in (b) as a function of q/m.

Fig. 5 shows the maximum deuteron energy versus laser energy for different target thicknesses. Each dot represents a laser shot and contains the information of the laser energy, maximum energy, and target thickness. From the study of CH targets, we expect that the maximum ion energy increases linearly with the laser energy. We apply this model to the data on deuteron acceleration with the wedge filter in place, even though it does not display the linear scaling law as clearly, especially for thin target shots with low laser energies, probably due to the reduction in signal from the $C^{6+}$ filter. Although the individual fits exhibit large uncertainties, upon comparing the maximum attainable energy (MeV per Joule of the laser energy) with the target thickness (in nanometers), we found that the thicker targets perform consistently better in terms of the deuteron acceleration (see fig 5b).

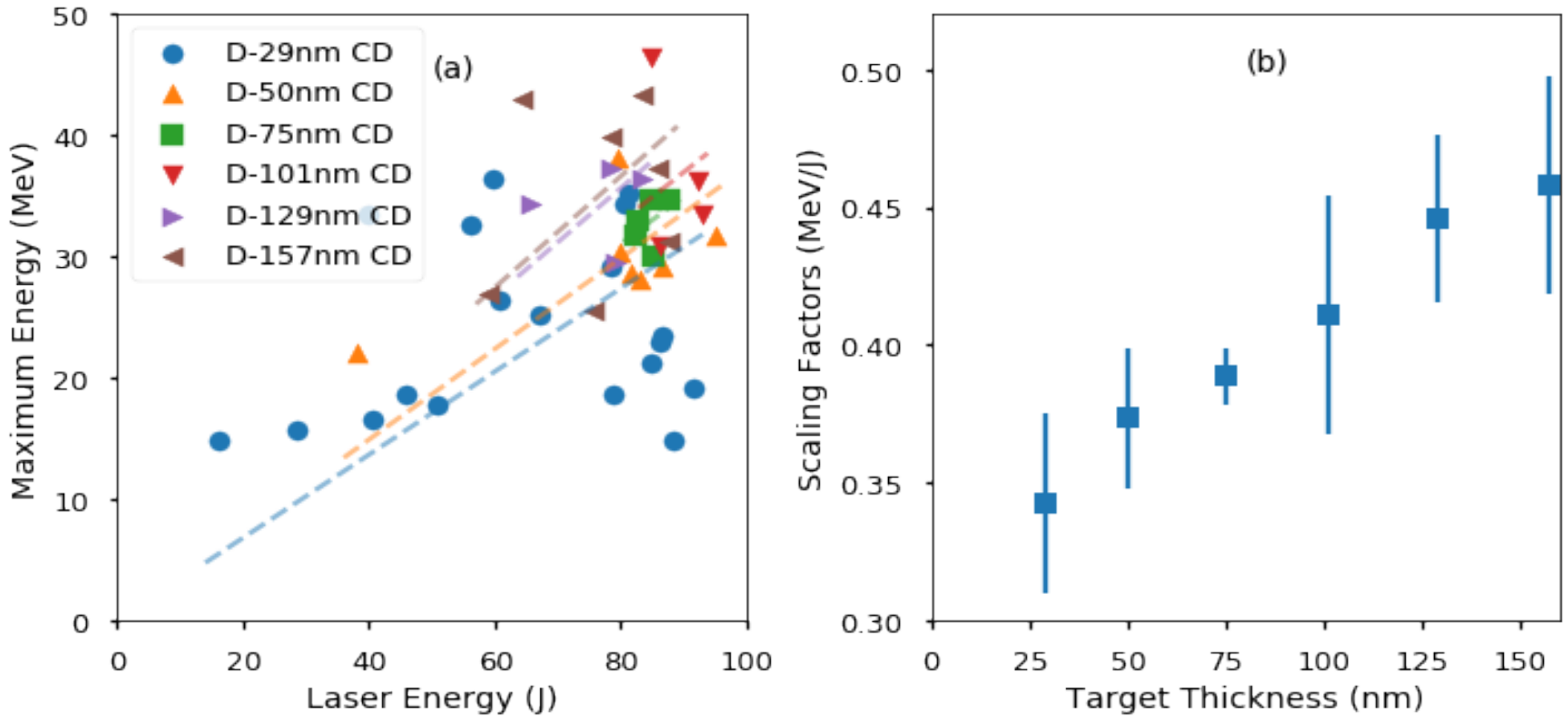

Figure 5 Maximum deuteron energy. (a) Maximum deuteron energy as a function of laser energy, separated by CD thickness. Dashed lines show linear fits. (b) Slope of linear fit as function of CD thickness (presumably error bar shows uncertainty)

In the second phase of the campaign, the copper converter was placed in the path of the deuteron beam generated by $CD_2$ targets of various thicknesses. Fig 6a shows the neutron spectra acquired from the n-TOF detector in the forward direction. The spectra were extracted by first subtracting the gamma-ray signal from the raw scintillation signal; this is accomplished by fitting the x-ray induced fluorescence of the plastic scintillator with an exponential decay model, as shown in Fig 7a. Then the neutron signals were converted from the time domain to energy domain ($dS/dt \rightarrow dS/dE*\alpha$, where $\alpha=dE/dt$). Finally, the neutron number per energy interval (dN/dE) was obtained by dividing by a theoretically calculated calibration curve (dN/dS) using the method described by O'Reilly [17].

Neutrons with energy up to 15 MeV were observed. Thicker $CD_2$ pitchers yielded neutron spectra with slightly higher maximum energies. The 15 MeV maximum is consistent with expectations from basic kinematic considerations. We start by noting that the maximum deuteron energy observed on 30-100nm $CD_2$ targets is typically 30-35 MeV (see figure 5) and that the deuteron breakup reaction is endothermic with Q=-2.2 MeV. The total kinetic energy of the (p+n) system after break-up should be around ~30MeV. Since the Cu nucleus is much heavier, most of this energy should be divided between the two nearly equal-mass product particles, namely proton and neutron. Therefore, we should expect the highest neutron energy to be around 15 MeV and thicker targets should yield higher neutron energy cutoffs because they produce deuterons with higher energies. Our observations are consistent with these assumptions.

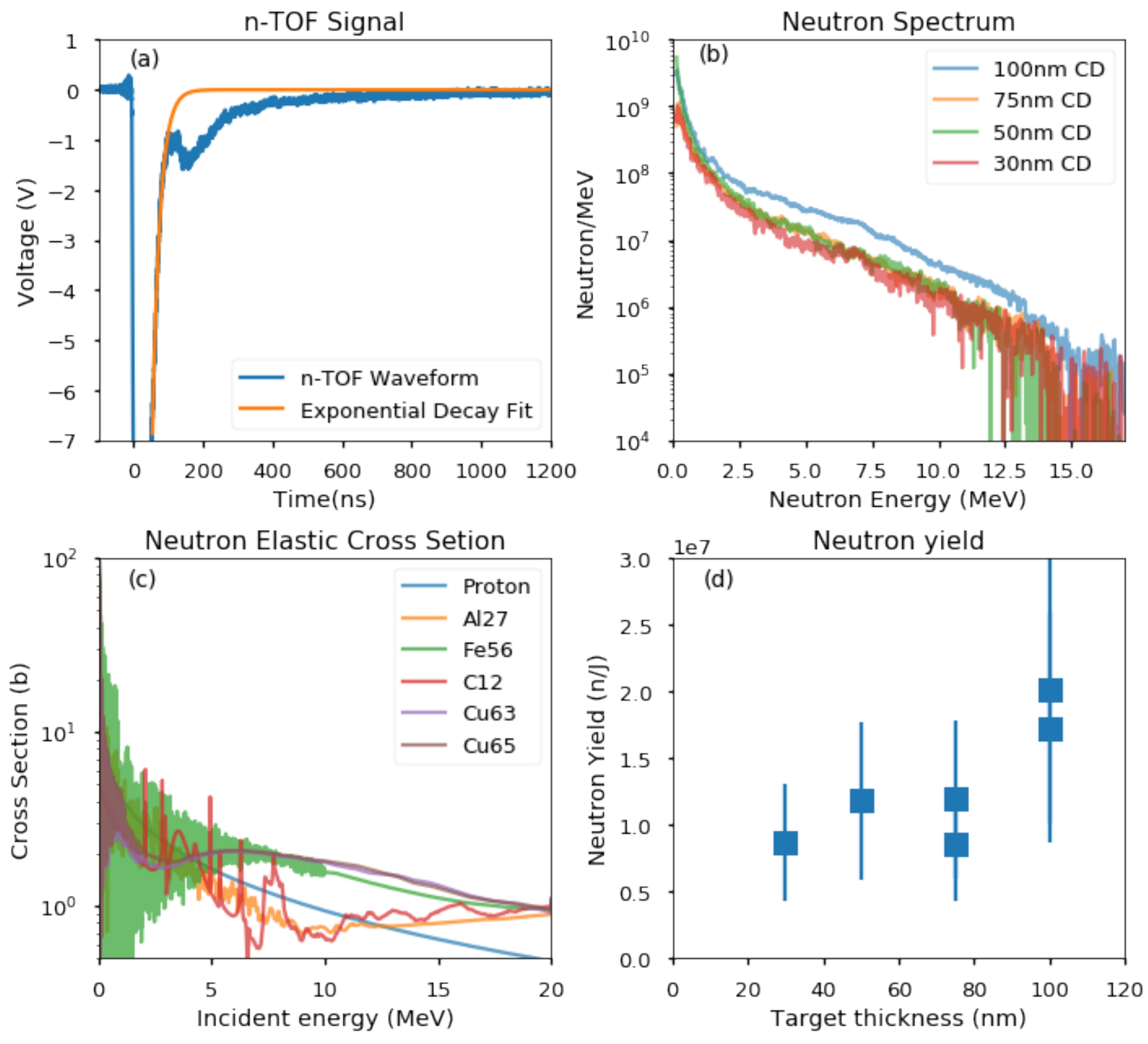

Figure 6 a) A typical n-TOF signal recorded at the forward direction. b) Neutron spectra calculated from the n-TOF signal. c) Neutron elastic cross section on common elements present in the laboratory. d) Neutron yield for CD target of different thicknesses. The total neutron yield is measured using bubble detector.

The high energy (>2 MeV) region of the neutron spectrum follows an exponential decay distribution, $dN/dE \simeq (N_0/0.4\ \text{MeV})\exp\,(-E/0.4\ MeV)$ with slope ~0.4/MeV. In contrast, the low energy region (< 2 MeV) follows a super exponential function. This super exponential distribution likely results from the environmental scattering, calibration and scintillation afterglow effects, which could occur at the target chamber, detectors, inside apparatus and the 10 cm thick plastic neutron radiation shield that is installed against the chamber wall and covers the angle from -30 to +30$^o$. The neutron radiation shield could not be removed, as it was required by governmental safety regulation. The neutron elastic scattering cross section on these elements is relatively flat for energies above 2 MeV but increases as the neutron energy decreases as shown in Fig. 6c. Some fractions of the high energy neutrons scatter during the transport and lose energy in the process. As they lose energy, the chance of being scattered increases, leading to further energy loss. Also, neutrons emitted toward the other directions could be scattered back towards the front, but the longer path length leads to delayed detection and thus identification as lower energy. All these contribute to the super exponential distribution at lower energy.

The total neutron yield per shot was determined from the bubble detector array distributed around the chamber walls at approximately 1.5 m from the copper converter. The bubble count in each detector was converted to dose (mrem) using the manufacturer-provided sensitivities (~20 bubbles/mrem), then to neutron fluence (n/cm$^2$) using the standard dose-to-fluence conversion for fast neutrons. The neutron flux per unit solid angle at each position was obtained as $\Phi(n/sr) = F \times r^2$, where F is the fluence and $r \approx 150$ cm. The neutron spectrum indicates that the majority of neutrons have energy at or below 1 MeV, meaning their mean free path in natural copper is <3.5 cm and therefore >75% scatter at least once in the converter. The forward-directed momentum of the neutrons inherited from the deuteron beam is substantially broadened but not made fully isotropic. Since neutrons in this energy range dominate the total number, we estimate the total yield by averaging $\Phi$ across detector positions and integrating over $2\pi$ sr for the forward hemisphere. Detectors behind the 10 cm plastic radiation shield (−30° to +30°) were excluded from the average due to neutron moderation by the shield.

Fig 6d shows that neutron yield increases with target thickness. This observation is explained with the help of the deuteron acceleration study, which showed that the maximum deuteron energy increases with target thickness. Since higher energy deuterons can travel deeper into the converter, they have more chances to react, which implies a positive correlation between the neutron yield and target thickness.

## Discussions and Conclusions

Even though the observed ion energy spectra are broadband and follow an exponential distribution, the ion acceleration observed in this experiment differs from TNSA in several observed characteristics. First, the trace found in 30 degrees is insignificant compared to the TPS signal at the laser direction. In contrast, a TNSA ion beam typically has a wide angular divergence with the strongest emission close to target normal direction. Second, the maximum ion energy scales linearly with laser intensity, whereas for TNSA it scales with the square root of the laser intensity [19][20]. Third, in this campaign, the other ion species were accelerated comparably to the protons even without in-situ cleaning of the target surface prior to firing the laser [21], whereas in TNSA protons are predominately accelerated due to their light q/m ratio. The observed ion acceleration also does not fit the radiation pressure acceleration prediction that maximum ion energy is inversely proportional to target thickness [3][6]; we found a positive correlation in our experiment.

To sum up, we presented a comprehensive analysis of a medium repetition rate laser-driven ion sources based on ultra-thin targets and used the accelerated deuterons to power a neutron source. We attribute the relatively high consistency of the collected data to the use of uniformly-mounted targets, precise target alignment techniques and the short laser firing cycle. Our higher shot count also strengthened the statistical analysis on the ion acceleration performance. The maximum energies of all ion species from both $CH_2$ and $CD_2$ targets were found to scale linearly with laser intensity with small uncertainties. The total energy fluences through the TPS pinhole scale approximately quadratically for all ion species. Comparing spectra from different ions and charge states, we found that the maximum ion energy is proportional to $q^2/m$, namely charge squared over mass. This scaling law is similar to the pondermotive force on ions, but we do not know of any models of ion acceleration exhibiting this scaling.

Using the deuteron ion source, we have generated a fast neutron source with exponential energy spectrum up to 15 MeV. Neutron yield $>10^7$ n/J was observed by shooting 100 nm thick $CD_2$ targets with a copper converter. The total yield duringt his campaign exceeded 10^9 neutrons per shot in selected cases. Our work represents the first ion-driven neutron production experiment performed with a multi-petawatt laser. All the neutron source properties are in good agreement with the deuteron beam spectral measurements, which confirms deuteron breakup as the primary source. This experiment demonstrates the advantages of collecting higher statistics in petawatt-class laser-plasma interactions, and its high yield of fast neutrons points to the advantages of thin targets with ultra-high contrast for ion-driven neutron sources, complementary to electron-beam-driven photoneutron sources recently demonstrated using laser wakefield acceleration [22].

Acknowledgement

Work performed under the auspices of the University of Texas at Austin. This work was supported by the Air Force Office of Scientific Research (FA9550-14-1-0045), by the U.S. Department of Energy, National Nuclear Security Administration, under Award Number DE-NA0004201, by the National Science Foundation under Grant Number NSF2108921; and by the Air Force Office of Scientific Research under Award Number FA9550-25-1-0286. This work was supported by the Institute for Basic Science under IBS-R038-D1. We would like to thank the CoReLS laser facility staff for their brilliant and unwavering support.

The data that supports the findings of this study are available from the corresponding author upon reasonable request.